\documentclass{article}

\usepackage[superscript]{cite}
\usepackage{graphicx}
\usepackage{hyperref}
\usepackage{url}

\newenvironment{acks}[1]%
{\subsection*{\normalsize\bfseries Acknowledgements}\noindent #1}

\newenvironment{contrib}[1]%
{\subsection*{\normalsize\bfseries Authors Contribution}\noindent #1}

\begin{document}


{ \large 

Title: \\[1mm] 
The Predictive Individual Effect for Survival Data

\vspace*{1cm}
Authors: \\[1mm]
Beat Neuenschwander$^a$, Satrajit Roychoudhury$^b$, Simon Wandel$^a$, \\ Kannan Natarajan$^b$, Emmanuel Zuber$^a$ \\[3mm]
$^a$ Novartis Pharma AG, Basel, Switzerland, 
$^b$ Pfizer Inc, New York, New York, USA


\vspace*{1cm}
Corresponding author:\\[1mm]
Satrajit Roychoudhury \\
Pfizer Inc, 235 E 42nd St,
New York, NY 10017 \\
email: satrajit.roychoudhury@pfizer.com \\




\vspace*{1cm}

{Key words:}\\[1mm]  Bayesian predictive inference, non-proportional hazards, patient-centric measure, rank preservation, survival gain, time-to-event endpoint
}

\newpage

{\large \bf Abstract}

\vspace*{5mm}

The call for patient-focused drug development is loud and clear, as expressed in the 21st Century Cures Act and in recent guidelines and initiatives of regulatory agencies. Among the factors contributing to modernized drug development and improved health-care activities are easily interpretable measures of clinical benefit.  In addition, special care is needed for cancer trials with time-to-event endpoints if the treatment effect is not constant over time.
We propose the {\em predictive individual effect} which is a patient-centric and tangible measure of clinical benefit under a wide variety of scenarios. It can be obtained by standard predictive calculations under a rank preservation assumption that has been used previously in trials with treatment switching. We discuss four recent Oncology trials that cover situations with proportional as well as non-proportional hazards (delayed treatment effect or crossing of survival curves). It is shown that the {\em predictive individual effect} offers valuable insights beyond p-values, estimates of hazard ratios or differences in median survival. Compared to standard statistical measures, the {\em predictive individual effect} is a direct, easily interpretable measure of clinical benefit. It facilitates communication among clinicians, patients, and other parties and should therefore be considered in addition to standard statistical results.

\newpage


\section*{Introduction}
The 21st Century Cures Act 
\cite{curesact2015} 
requests change, which includes topics like modernizing clinical trials, incorporating patient perspectives into medical research and regulatory processes, 
precision/personalized medicine, and digital health care. 

A more patient-centric perspective calls for measures of clinical benefit that are easy to understand and have a clinically meaningful interpretation. In Oncology, for example, the hazard ratio has come under scrutiny recently because it lacks interpretability in settings with non-proportional hazards. Recent clinical trials with immuno-oncology agents have shown clear deviations from the proportional hazards assumption, such as delayed separation or crossing of survival curves  
\cite{Motzer2015,Ribas2013,Wolchok2010,Larkin2015, Postow2015,Fehrenbacher2016,fer2016nfr}. For such scenarios, alternative measures of clinical benefit are needed to describe the time-dependent treatment effects.

The estimand framework contributes to a better understanding of treatment benefits and aims for improved decision making 
\cite{ich2019aoe}. While it puts strong attention to the handling of intercurrent events, it also aims at more specific clinically 
meaningful measures that are easily understandable for patients, physicians, regulators, payers, and the scientific community. Easily interpretable measures enhance the communication between physicians and patients, which contributes to patient-focused drug development 
\cite{fda2020pfdd,ema2020pfdd,eps2005mpcc}.

\subsection*{Treatment effects for time-to-event trials}

Traditional statistical approaches to comparing treatments in clinical trials use hypothesis testing and estimation methods. For cancer trials, time-to-event (e.g., disease progression, death) is often the primary endpoint. The three main summary measures are the p-value from the log-rank test, the estimated hazard ratio from the Cox proportional hazards model, and difference of median survival times. However, they have well-known limitations.

First, dichotomizing trial results with the statistical significance criterion based on p-values is problematic. There is an increasing understanding among the medical-statistical community that decision making 
goes far beyond a simple consideration of whether a study reached statistical significance 
\cite{gib2021rop,goo1999pval,ham2021edi,nea2018ts,oco2021mm,
was2016asa,roy2018dcd}.
This implies the need for effect measures that are intuitive and simple to interpret, which improves communication between all parties.

Second, hazard ratios (or relative risk measures in general) 
are often misinterpreted \cite{spi2019aos} and do not answer questions of direct interest (e.g., gain in overall survival) in a straightforward manner. Moreover, departures from the proportional hazards assumption hamper the interpretation of the hazard ratio for the treatment effect 
\cite{aaa2015dca,her2010hohr}.

Third, while differences in median survival times are often useful, they may not capture the treatment effect sufficiently well (e.g. for crossing survival curves), 
their estimation may be unstable (when only a few patients are at risk) or may not even be possible, 
for example if the percentage of patients with an event is too low for one of the treatment arms.

Selecting an appropriate method for summarizing the treatment effect depends on the true underlying survival curves (which are not known precisely), and 
interpretation of results can be controversial. Given these difficulties, Freidlin and Korn
\cite{fre2019mfa} 
rightly ask the question "Ready for the primary analysis?" and discuss various testing and estimation methods for the various treatment effects; see also Lin et al., Roychoudhury et al. and Uno et al.
\cite{lin2020aam,roy2021rda,uno2014mbhr}  
for inferential options and recommendations. In addition to the hazard ratio and difference in medians, alternatives include the difference of restricted mean survival times and the difference of survival probabilities for selected milestone dates (e.g., one year survival) as well as alternative survival models \cite{kot2001ftmm, kay2002aftm}. Deciding on an appropriate measure may not always be easy, and more generally applicable measures would be advantageous.

Irrespective of the selected method, one should always keep in mind that testing and estimation approaches are reductionist in nature. While useful, their results will often provide only partial answers to what one would ideally like to know. Testing addresses the question of whether the survival curves differ, not in what respect they differ. On the other hand, any of the summary measures (e.g., the hazard ratio, difference in medians) only targets a specific aspect of the treatment effect, which may not reflect all relevant differences nor how they translate at the individual patient level.

Decision making is always subjective to some degree because stakeholders may interpret the evidence differently. Hence, it is important to present the available information in different ways. In addition to the p-value, the Kaplan-Meier survival curves and other relevant information should always be used for decision making.

We propose going a step further and considering an individual's predicted survival gain, the {\em predictive individual effect}, when choosing one treatment over the other.

\section*{Methods}

What additional information may be useful for decision makers and patients? We will try to answer the following question: 

\begin{quote}
What is the difference in survival {\bf for a new patient}, would he/she be treated with the test or control treatment?
\end{quote}
Since a new patient is considered, both survival times, $X$ for control and $Y$ for test, are unknown. So they must be predicted based on the available data, and we therefore refer to the difference of survival times as the {\em predictive individual effect}, that is
\[
\mbox{eff}(X,Y) = Y-X
\]
The predictive individual effect answers a clinically relevant question, is easily interpretable, and has a direct causal interpretation 
\cite{rub2005ciu,her2020whatif}. 
Here, the effect is defined as the difference of survival times, but there may be situations where the ratio $(Y/X)$ may be of interest.
The predictive individual effect will never be known precisely. Its uncertainty is expressed by its probability distribution derived from observed control and test data $X_{obs}$ and $Y_{obs}$
\[
pr(Y-X \vert X_{obs},Y_{obs})
\]
For survival data, the distribution of the predictive individual effect captures an individual's predictive survival gain (or loss). The distribution can be summarized using standard summaries such as the mean or median, the 95\% interval, the probability that an individual's survival under the test treatment is longer than under control, or the probability that it is longer by a clinically relevant margin (e.g. three months).   We recommend communicating the full distribution of the predictive individual effect to decision makers, as shown in the Examples Section.

Of note, the probability that survival is longer under test than under control has been discussed previously by various authors \cite{aci2006pi,buy2008rhr,mos2008rhr}. 
Their perspective differs from ours in that they consider survival times from different individuals rather than the same individual, which has been discussed controversially 
\cite{aci2006ar,sen2006pi}.

The simplest version of the predictive individual effect quantifies the survival gain in the absence of information related to factors that  influence the treatment effect. Refined versions that include such factors are possible.

Figure 1 summarizes the predictive individual effect, showing predicted test and control survival times for four new subjects based on trial data $X_{obs}$ (control) and $Y_{obs}$ (test).

\begin{center}
(Figure 1 around here)
\end{center}

\subsection*{Statistical aspects}

From a statistical standpoint, obtaining the distribution of the predictive individual effect $Y-X$ must address three issues. 

First, the survival curves for test and control are not known precisely and must be estimated from the available data. We use a parametric approach and select the distributions for test and control from the flexible family of piecewise exponential distributions using the deviance information criterion 
\cite{spi2002bmm} 
and account for parameter uncertainty of the underlying hazards in a Bayesian way using weakly informative prior distributions (see on-line appendix for details).

Second, the prediction of test and control times $Y$ and $X$ must account for predictive uncertainty. This refers to the fact that, even if the survival distributions for control and test were known precisely, there remains the uncertainty about a new patient's survival time. For piecewise exponential distributions, the predictive distributions for $X$ and $Y$ can be obtained easily using standard Bayesian computations. They build on a statistical model $pr(Y \vert \theta)$, from which the predictive distribution
for a new observation $Y_{n+1}$ after observing $Y_1,\ldots,Y_n$ follows as
\[
pr(Y_{n+1} \vert Y_1,\ldots,Y_n) = \int pr(Y_{n+1} \vert \theta)
pr(\theta \vert Y_1,\ldots,Y_n) d\theta
\]
which shows the two main sources of uncertainty that drive predictions, that is, sampling uncetainty $pr(Y_{n+1} \vert \theta)$ and parameter uncertainty $pr(\theta \vert Y_1,\ldots,Y_n)$.
Here, weakly-informative Gamma priors are used for the hazard parameters of 
the piecewise exponential model. Alternatives to the Bayesian implementation may be envisaged, such as the maximum likelihood predictive density\cite{lej1982mlpd}, fiducial or nonparametric approach.

The third issue is more fundamental: the survival distributions for test and control do not suffice to determine the predictive individual effect because X and Y refer to the same individual.  Thus, an assumption about the dependency of $X$ and $Y$ is needed. We use a model introduced by Lehmann 
\cite{leh1974nsm} 
and investigated by Doksum 
\cite{dok1974epp} 
(see on-line appendix). One version of their model is equivalent to the rank preservation property underlying rank preserving structural failure time (RPSFT) models, which have been used to adjust analyses for trials with treatment switching 
\cite{rob1991cnc}.

Simply put, rank preservation means that for any two subjects in the population of interest (say, subject 1 and 2), if subject 1 would fail before subject 2 under the control treatment, then subject 1 would fail before subject 2 under the test treatment. 

Rank preservation may be questionable, in particular
in the presence of a predictive biomarker. For example,
a significant survival benefit for nivolumab vs. docetaxel has been seen in a recent study of metastatic non-squamous NSCLC patients with the PD-L1 biomarker expressed as compared to those without 
\cite{bor2015nvd}.
In such a situation, the predictive individual effect should be assessed separately in each biomarker group (assuming rank preservation within each group rather than for the full population); see on-line appendix for this example.

\section*{Results}

We now look at the predictive individual effect for four recent cancer trials. The examples include scenarios with delayed effect, crossing survival curves, and proportional hazards. The data sets were reconstructed from the published Kaplan-Meier plots 
\cite{Guyot2012}, so some of the standard inferential summaries presented here deviate slightly from the ones in the original publications.

Our purpose is to demonstrate the utility of the predictive individual effect. We have no intention to judge the clinical activity of the treatments involved or any related regulatory decisions. 
For each example, results will be summarized with the following three plots:
\begin{enumerate}
\item[A.]
	Kaplan-Meier survival curves for the control and test treatment. These plots should be an integral part of any decisions related to treatment selections for patients.
\item[B.]	
	The cumulative distribution of the survival gain or loss for test compared to control, that is $pr(Y-X>d)$ for positive values of d (survival gain) and $pr(Y-X<d)$ for negative values of d (survival loss). Examples comprise the probabilities that an individual's survival is longer under the test treatment than under the control treatment $pr(Y>X)$, that it is at least 3 months longer $pr(Y>X+3)$, or that it is shorter $pr(Y<X)$.
\item[C.]	The conditional distribution of the survival gain $Y-X$ (summarized by the median and 95\%-interval) for selected values of the control survival time $(X=c)$, reflecting different levels of prognosis. This provides further insights into the survival gain for subjects with a bad (small $c$) or good (large $c$) prognosis.
\end{enumerate}

\subsection*{1. CheckMate 141}

The first example (CheckMate 141) is a randomized clinical trial in patients with recurrent or metastatic squamous cell carcinoma of the head and neck after platinum chemotherapy \cite{fer2016nfr}.This population has poor prognosis and limited therapeutic options. The study compared standard therapy (methotrexate, docetaxel, certuximab) to nivolumab, an anti-programmed death 1 (PD-1) monoclonal antibody. 
The Kaplan-Meier curves (Figure 2A) show a promising but delayed effect of nivolumab after approximately four months. The estimated overall hazard ratio for death is 0.70 in favor of nivolumab, with corresponding log-rank test p-value 0.0034. Median survival times are 7.5 months (95\% CI: 5.5-9.1 months) for nivolumab and 5.1 months (95\% CI: 4.0-6.0 months) for standard therapy.

Figure 2B shows the distribution of the predictive individual effect, represented as the cumulative survival gain. A patient's chance for a longer survival (survival gain $>0$ months) under nivolumab compared to chemotherapy is 82\%. Given the poor prognosis of the disease, the potential survival gain appears considerable: for example, the chance of surviving at least three months longer under nivolumab is approximately 47\%, and the chance of surviving at least six months longer is 27\%.

In addition, Figure 2C shows that while patients with an overall poor prognosis (as manifested by control survival times of one to three months) have no survival gain, the survival gain increases with improved prognosis; for example, a patient with control survival time of six months has a predicted survival gain (95\%-interval) of approximately 4.6 (1.6-8.6) months.	

\begin{center}
(Figure 2 around here)
\end{center}

\subsection*{2. IMvigor211}

The second example (IMvigor211) is a trial for patients with locally advanced or metastatic urothelial carcinoma after progression with platinum-based chemotherapy, comparing atezolizumab (anti-programmed death ligand 1 [PD-L1]) to chemotherapy 
\cite{pow2017avc}. 
Here, we consider the overall survival endpoint for the patients in the intention-to-treat population and with $>1\%$ tumor-infiltrating immune cells (PD-L1 expression: IC1/2/3). 
The median survival times are 8.9 (95\% CI: 8.2-10.9 months) for atezolimab and 8.2 (95\% CI: 7.4-9.5 months) for chemotherapy, showing a modest difference of 0.7 months. The Kaplan-Meier curves (Figure 3A) show an early crossing at four to five months and then a survival benefit for the atezolizumab arm. The stratified Cox regression analysis is not significant (HR=0.87, 95\% CI: 0.71-1.05).

The distribution of the predictive individual effect shows that a patient's chance of longer survival under atezolizumab compared to chemotherapy is 70\% (Figure 3B). Moreover, the probabilities of surviving three or six months longer under atezolizumab are 32\% or 18\%. The traditional summary measures (log-rank p-value and HR) alone fail to capture this potentially important benefit for atezolizumab in advanced or metastatic urothelial carcinoma.

Finally, Figure 3C shows that atezolizumab becomes beneficial for patients with an anticipated survival prognosis of 10 or more months under chemotherapy.

\begin{center}
(Figure 3 around here)
\end{center}

\subsection*{3. NCIC CTG PA.3}

The third example (NCIC CTG PA.3) is a trial in patients with advanced pancreatic cancer, comparing gemcitabine with the combination erlotinib plus gemcitabine \cite{moo2007epg}. The primary endpoint was overall survival.
The Kaplan-Meier curves (Figure 4A) show separation. The study revealed a significant effect in favor of the combination therapy (p-value = 0.023), with an estimated hazard ratio of 0.82 (95\% CI: 0.69-0.99). Despite the significant result, the treatment benefit appears modest: for example, the estimated median survival times are 6.2 and 5.9 months for the combination and control, respectively.

In this example, a patient's chance of surviving longer under the combination therapy is high (91\%) (Figure 4B). However, despite the statistically significant finding, the chance of a clinically relevant survival gain decreases quickly; for example, the probability to survive at least one or two months longer under the combination are 32\% and 15\%, respectively.

The predicted survival gain as a function of the survival times under monotherapy (Figure 4C) is modest (approximately half to one month), for the vast majority of patients (control survival times of up to 10 months), which raises further doubt about the benefit of the 
combination therapy.

\begin{center}
(Figure 4 around here)
\end{center}

\subsection*{4. ASPIRE}

The fourth example (ASPIRE) is a trial in patients with relapsed or refractory multiple myeloma, comparing carfilzomib, lenalidomide, and dexamethasone (KRd) versus lenalidomide plus dexamethasone (Rd) 
\cite{sie2018aspire}. 
Median survival was 48.3 months (95\% CI: 42.4 to 52.8 months) for KRd versus 40.4 months (95\% CI: 33.6 to 44.4 months) for Rd. The hazard ratio was 0.80 (95\% CI: 0.67 to 0.95), with p-value 0.0047.
The Kaplan-Meier curves show clear separation (Figure 5A). 

In contrast to the third example, however, the statistically significant result translates into a clear clinical benefit. The chance of surviving longer under KRd is 97\%, and the chance of having a four, eight, and 12 months longer survival is 79\%, 50\%, and 26\%, respectively (Figure 5B).

The conditional survival gains (Figure 5C) are substantial although less pronounced for patients with poor prognosis.

\begin{center}
(Figure 5 around here)
\end{center}

\section*{Discussion}

Table 1 summarizes the distributions of the predictive individual effect for the four examples. The probabilities of a longer survival for the test treatments are 82\%, 70\%, 91\%, 97\%. 

As discussed, these percentages alone are of limited value. For a better understanding of the clinical benefit, the full distribution of the predictive individual effect should be considered. The summaries in Table 1 show that even though the chance of
a survival gain for the combination in NCIC CTG PA.3 trial (example 3) is higher than for the test treatments in examples 1 and 2, the survival gain is much smaller.

\begin{center}
(Table 1 around here)
\end{center}

Further details regarding the technical implementation of PIE for four examples are available at \url{https://github.com/roychs04/PIE}. It contains the R-library clinpredict, a technical Appendix and example code.

\section*{Conclusion}

Making informed decisions in the development and use of medical products is a challenging undertaking. Due to the ever increasing complexity of modern drug development, with a growing demand for meaningful information from patients and many more stakeholders taking part in the decision processes, the imperative for innovation and improvements at all levels is clear. Sponsors, regulators, payers, physicians, and patients must be aware of the benefits and risks of various treatment options. Even if they are aware, they have different perspectives, which can lead to a different understanding of clinical benefits or risks.

Answers to clinically relevant questions comprise statistical summaries for the efficacy and safety of a product. The most prominent measure to compare treatments is the p-value. However, extrapolating or translating p-values and other statistical summaries into clinically relevant metrics can be difficult. The need for better communication among all parties has been and still is recognized as important 
\cite{agg2018sl,bra2013cced,cal2016pcc,cam2017mtjn,
che2001nfc,mac2020mgv}. 

The predictive individual effect helps answering the question about a patient's survival gain, which is clinically relevant, is easy to understand for all parties, and has a causal interpretation. It not only fits the estimand framework
\cite{ich2019aoe,gel2020pmci}, 
it also aligns with the predictive paradigm introduced by de Finetti 
\cite{def1974top} 
and Geisser 
\cite{gei1993pii}, which differs conceptually from standard hypothesis testing or estimation approaches.

For survival data, the predictive individual effect seems particularly helpful for nonproportional hazards settings, when Kaplan-Meier plots reveal delayed treatment effects or crossing survival curves as seen in recent immuno-oncology trials. The traditional summary measures fail to give a clear picture of the treatment effect   
\cite{buy2020atb,roy2021rda}. 
In such cases, the predictive individual effect offers further insights and an intuitive interpretation of treatment benefits in relation to the overall prognosis of patients, which is relevant to patients and health care professionals. As seen in Examples 3 and 4, however, the predictive 
individual effect is also useful for cases where the proportional hazards assumption is sensible. More generally, it may also be of interest beyond time-to-event endpoints.

Here, our focus has been on the simplest version of the predictive individual effect, which does not use individual patient characteristics that may be informative about a patient's survival prognosis. While we have looked at the survival gain $Y-X$ for different anticipated survival times under control, implicitly reflecting a patient's prognosis, this does not explicitly acknowledge factors that influence the treatment effect. Indeed, once such factors are known, the predictive individual effect can then be determined for subgroups defined by these factors. We refer to the on-line appendix for an example with a predictive biomarker.

Our aim here was to introduce the basic idea of the predictive individual effect. Areas for future work include continuous biomarkers, other endpoints, weaker versions of the rank preservation assumption, alternative predictive approaches 
(non-Bayesian, non-parametric), and simulation studies to compare different approaches. 

In summary, we think the predictive individual effect supplements the arsenal of tools that help implementing the demands expressed in the 21st Century Cures Act and other proposals for patient-centric activities 
\cite{curesact2015,ema2020pfdd,fda2020pfdd}, 
supporting regulatory decision making, and improving communication among stakeholders.

\begin{acks}
	We would like to thank the referees, associate editor and editor for their insightful comments which greatly helped improve the manuscript.  
\end{acks}

\begin{contrib}
	BN and SR contributed to conceptualization, data collection, analysis, and writing. SW, KN, and EZ contributed to the writing and editing of this paper.
\end{contrib}

\bibliographystyle{vancouver}
\bibliography{benBiblio}

\newpage

\section*{Table}

\begin{table}[h]
\small\sf\centering
  \caption{Summaries of the predictive individual effect (survival gain) distribution for test vs. control $(Y-X)$ for the four examples}
\begin{tabular}{lcccc}
                      & mean & median & 95\%-int.   & $pr(Y>X)$ \\ 
CM141                 &  4.1 & 2.6  & ( -0.6, 18 )  & 0.82          \\
IMvigor211            & 3.1  & 1.0  & ( -1.0, 19 )  & 0.70          \\
NCIC CTG              & 1.2  & 0.6  & ( -0.3, 7.2 ) & 0.91          \\
ASPIRE                & 9.5  & 8.0  & ( -1.7, 32 )  & 0.97          
\end{tabular}
\label{tab1}
\end{table}

\newpage

\section*{Figures}

Figure 1: Data from actual trial and predictions for four new subjects

\vspace*{1cm}

\begin{figure}[!h]
\begin{center}
 \includegraphics[width=0.7\textwidth]{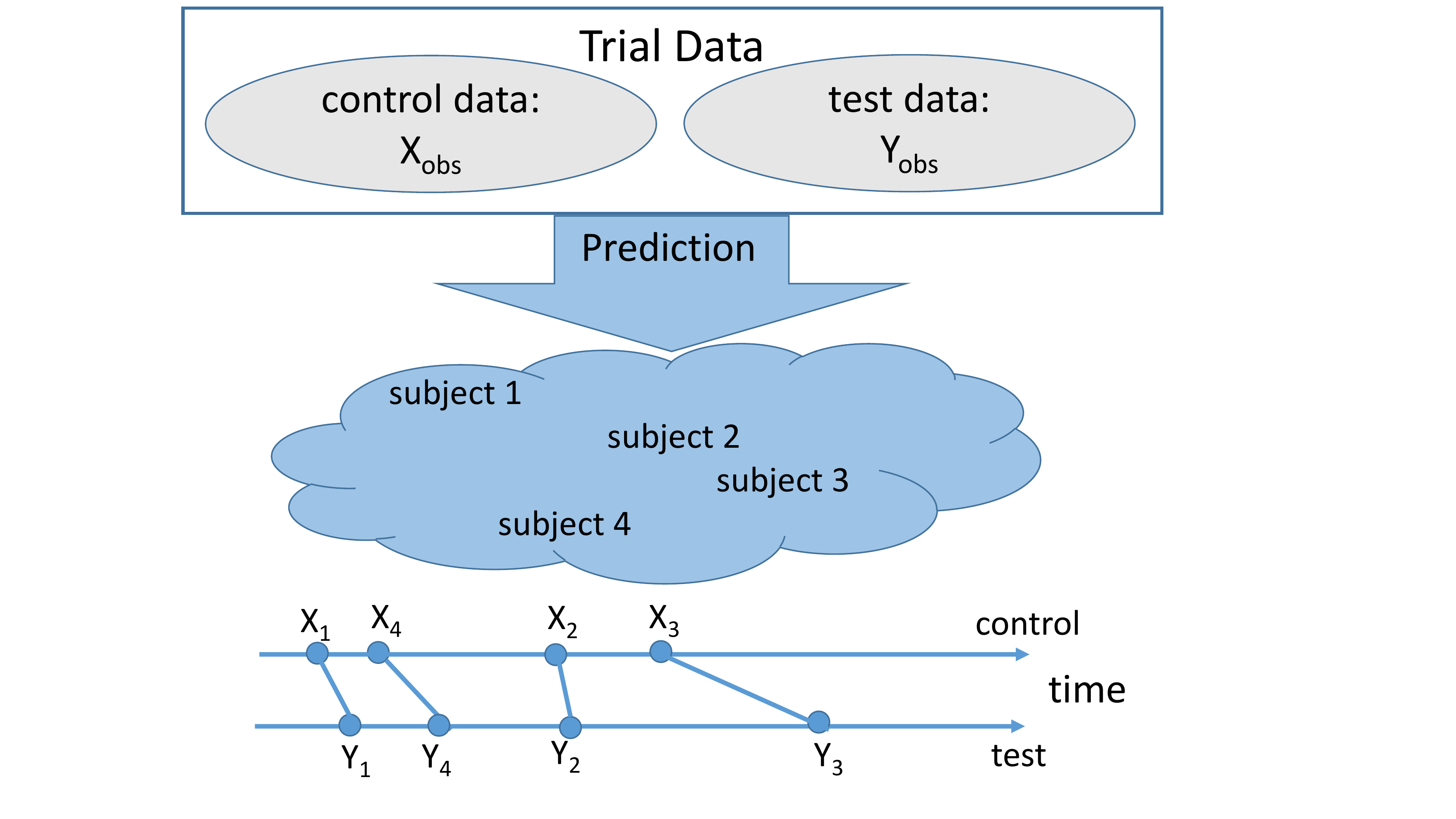}
\end{center}
\end{figure}

\newpage

Figure 2: CheckMate 141 trial: Kaplan-Meier curves, cumulative predictive individual effect (survival gain) distribution, and 95\% prediction intervals for survival gains conditional on control survival times

 \begin{figure}
 \includegraphics[width=0.48\textwidth]{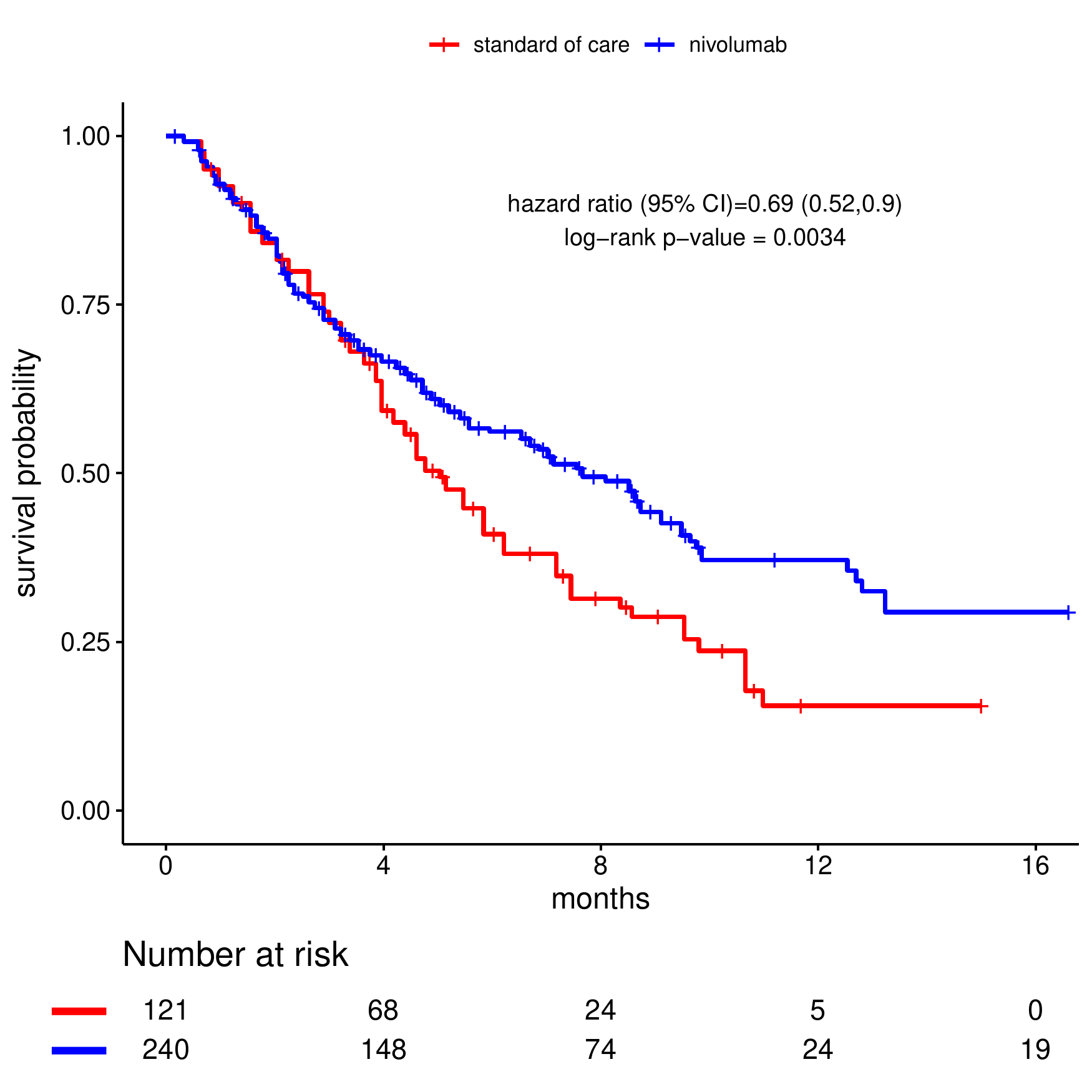}
 \includegraphics[width=0.48\textwidth]{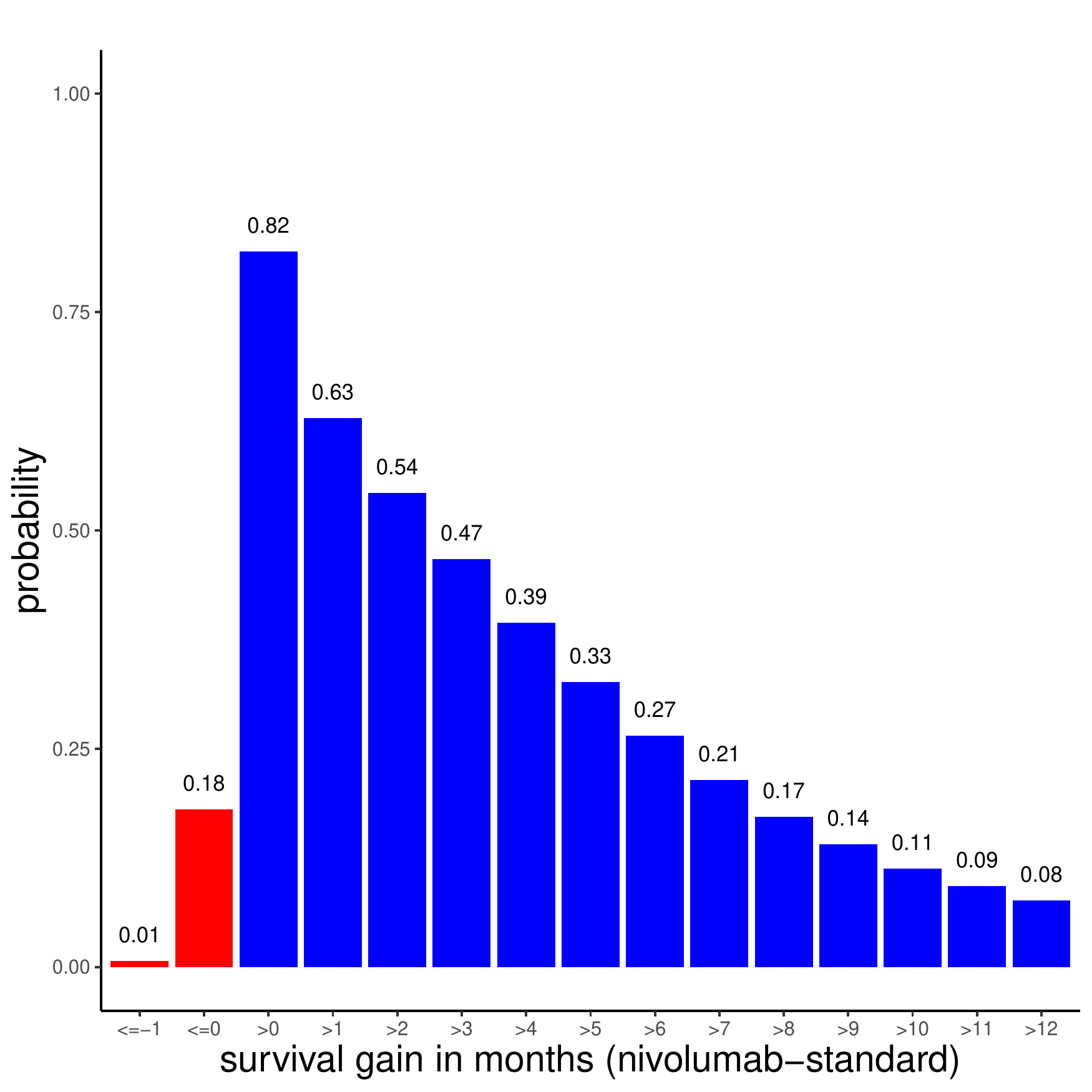}
 \includegraphics[width=0.48\textwidth]{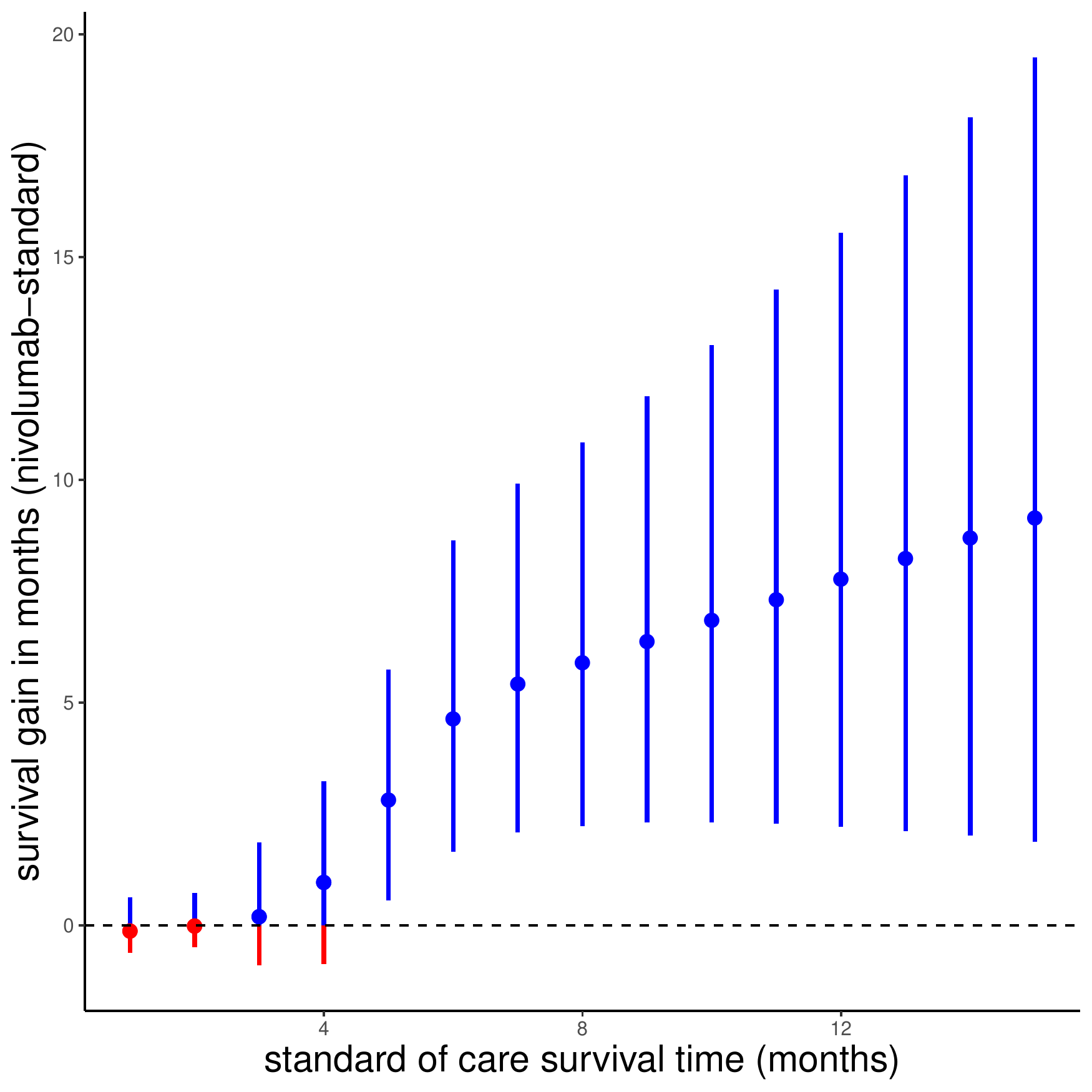}
\end{figure}

\newpage

Figure 3: IMvigor211 trial: Kaplan-Meier curves, cumulative predictive individual effect (survival gain) distribution, and 95\% prediction intervals for survival gains conditional on control survival times

 \begin{figure}
 \includegraphics[width=0.48\textwidth]{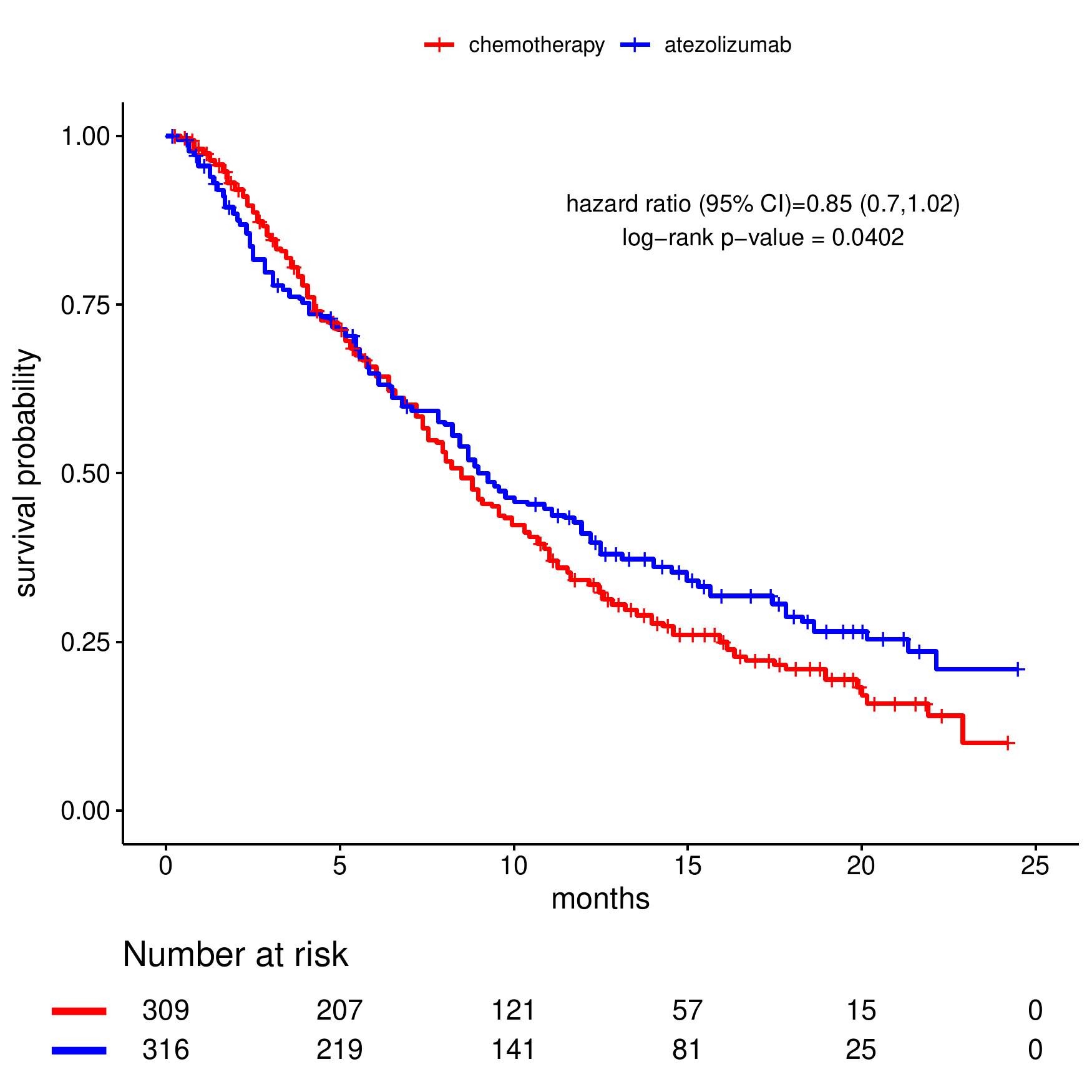}
 \includegraphics[width=0.48\textwidth]{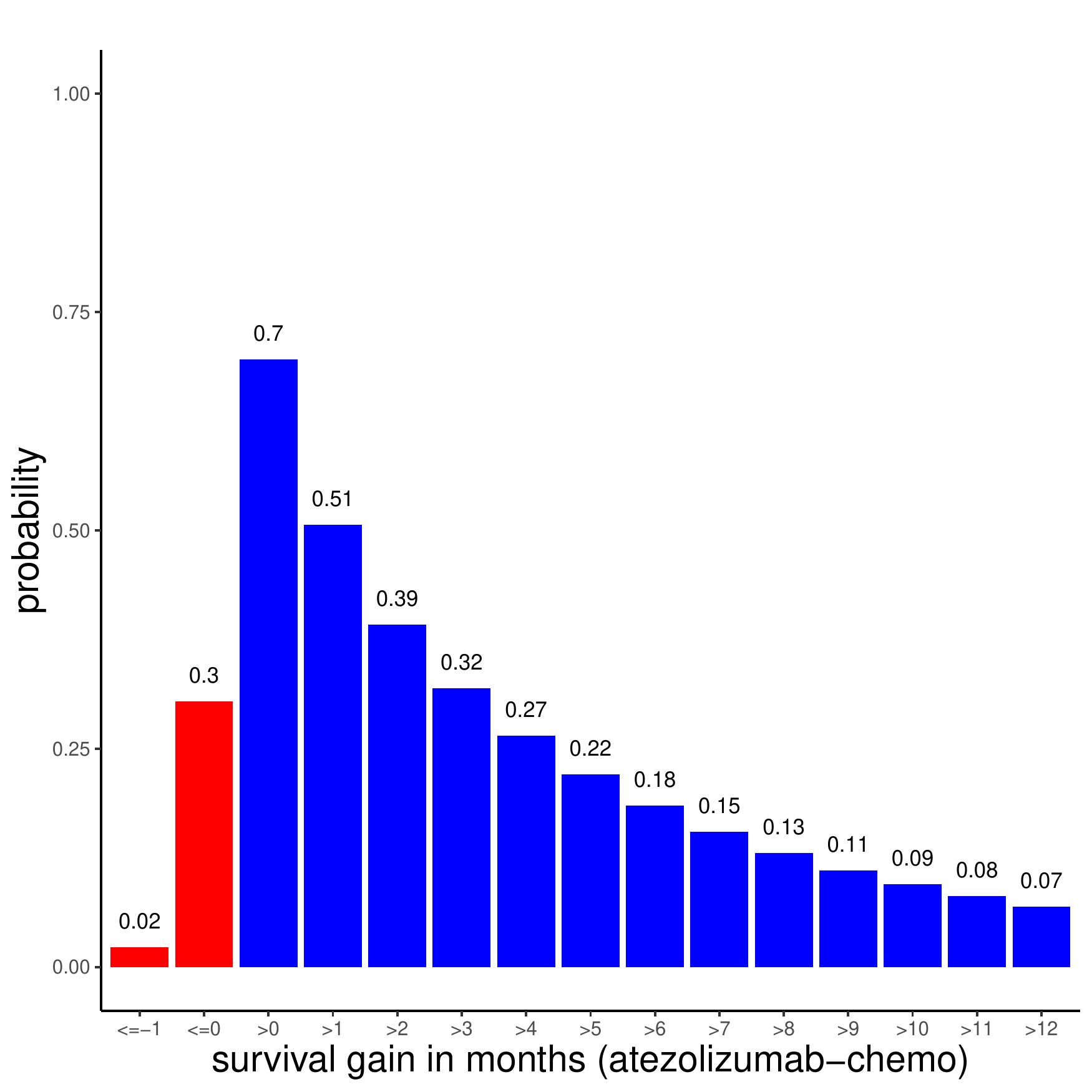}
 \includegraphics[width=0.48\textwidth]{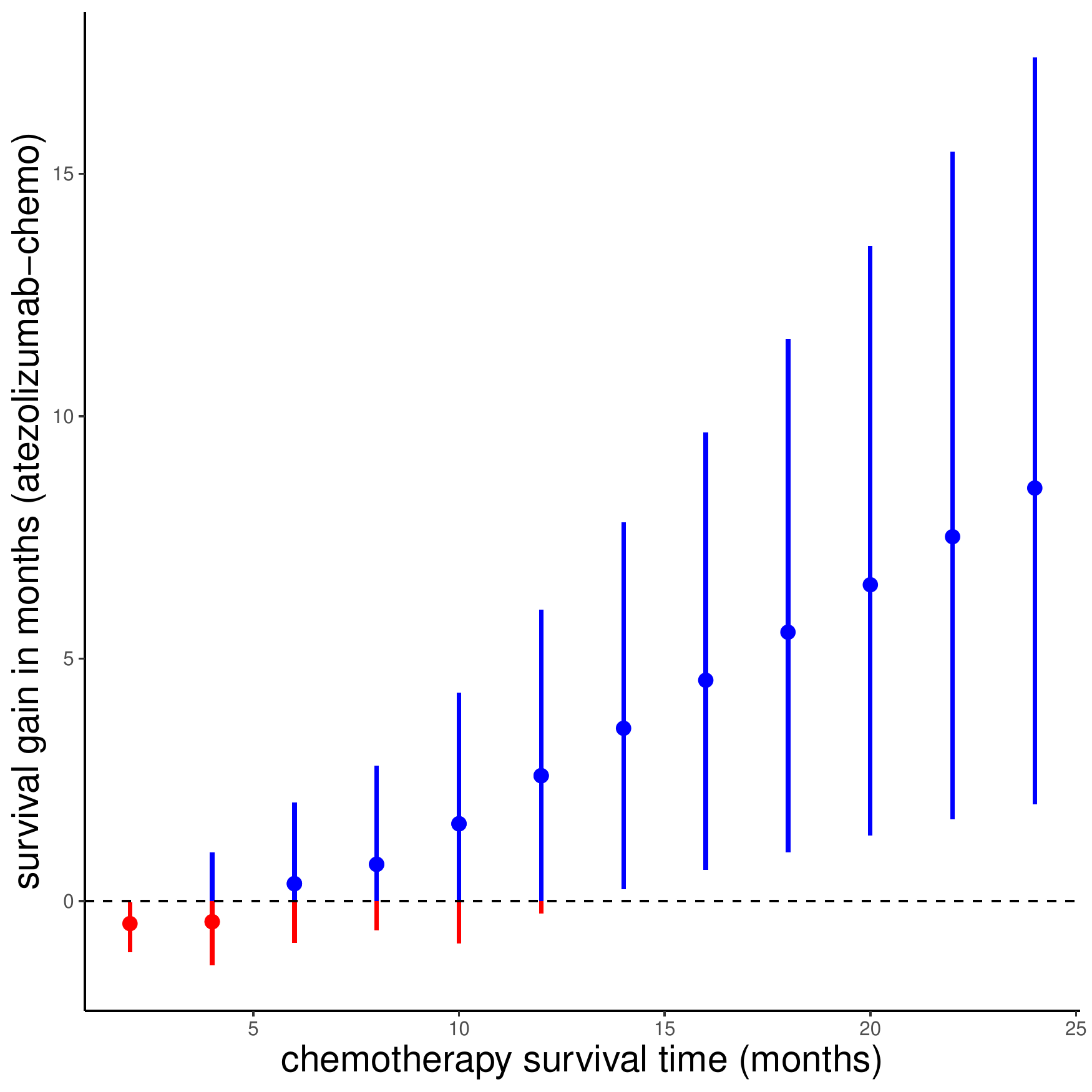}
\end{figure}

\newpage

Figure 4: NCIC CTG PA.3 trial: Kaplan-Meier curves, cumulative predictive individual effect (survival gain) distribution, and 95\% prediction intervals for survival gains conditional on control survival times

 \begin{figure}
 \includegraphics[width=0.48\textwidth]{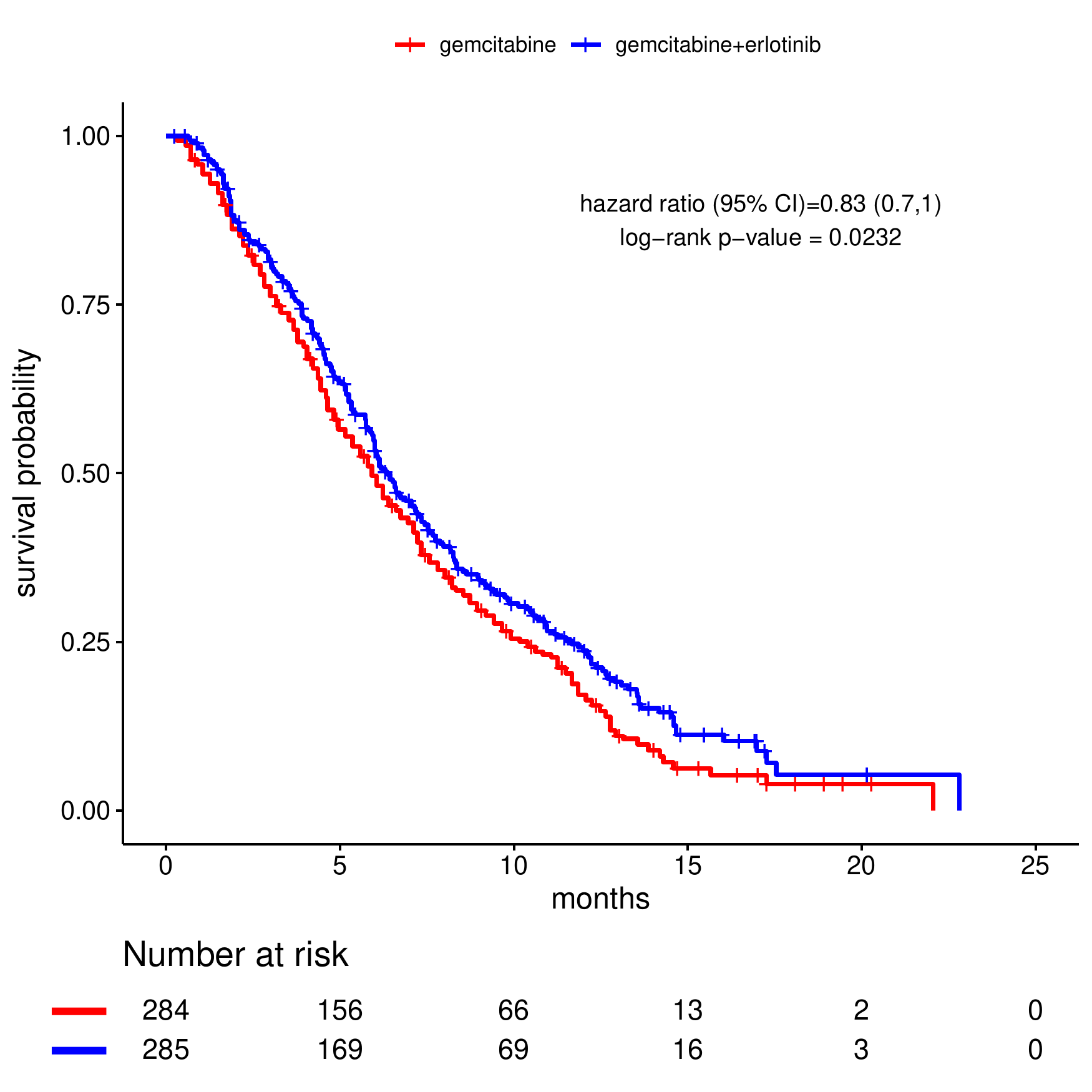}
 \includegraphics[width=0.48\textwidth]{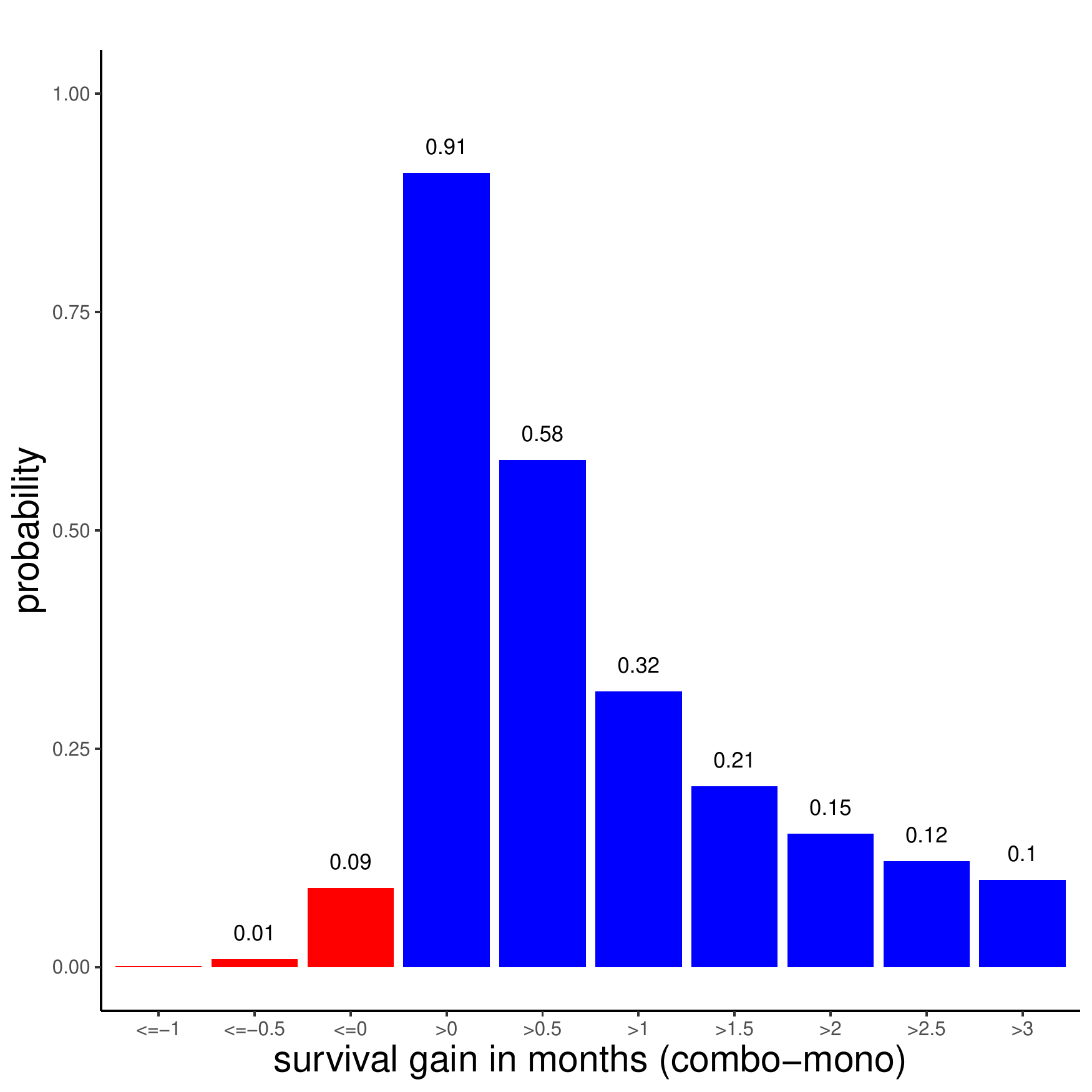}
 \includegraphics[width=0.48\textwidth]{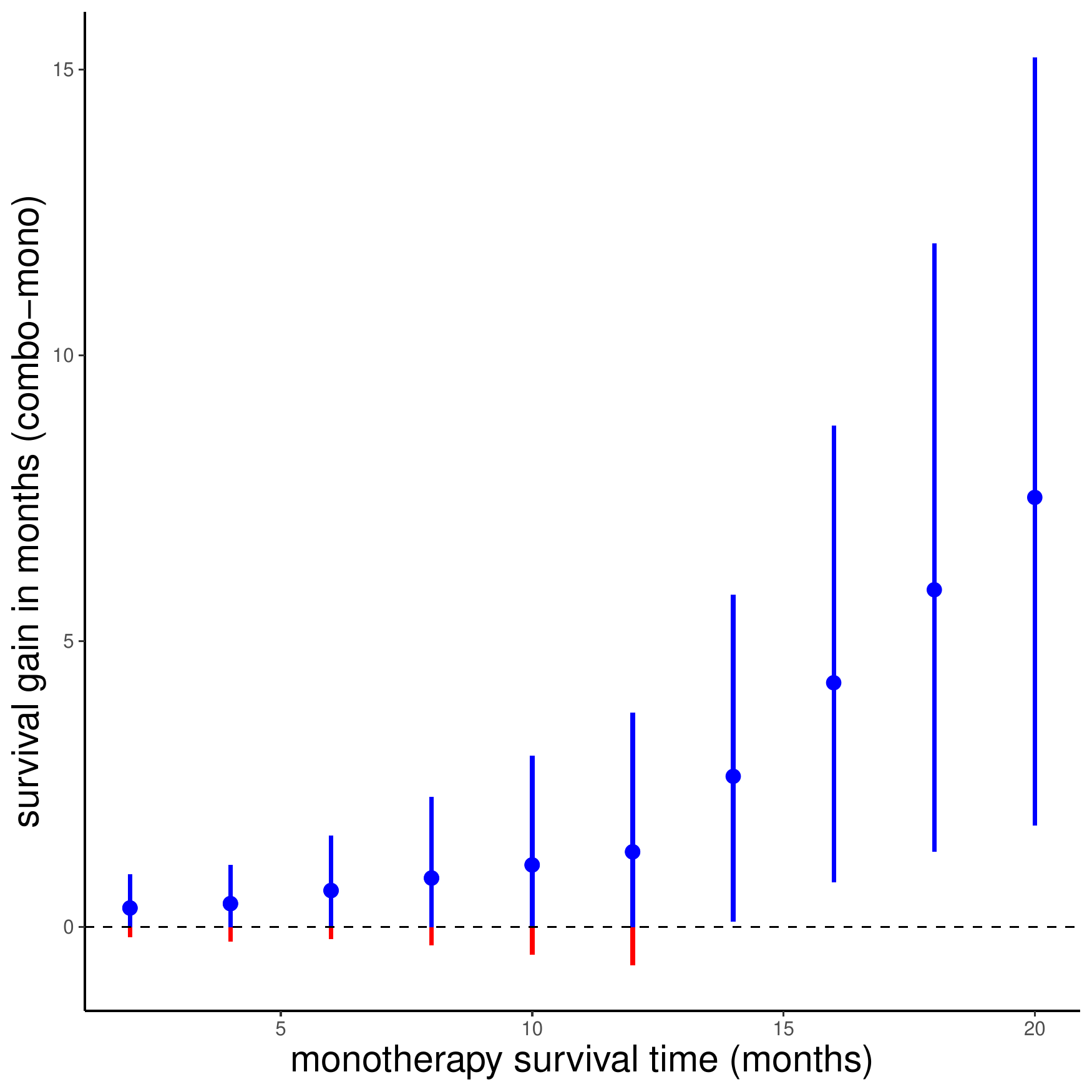}
\end{figure}

\newpage

Figure 5: ASPIRE trial: Kaplan-Meier curves, cumulative predictive individual effect (survival gain) distribution, and 95\% prediction intervals for survival gains conditional on control survival times

 \begin{figure}
 \includegraphics[width=0.48\textwidth]{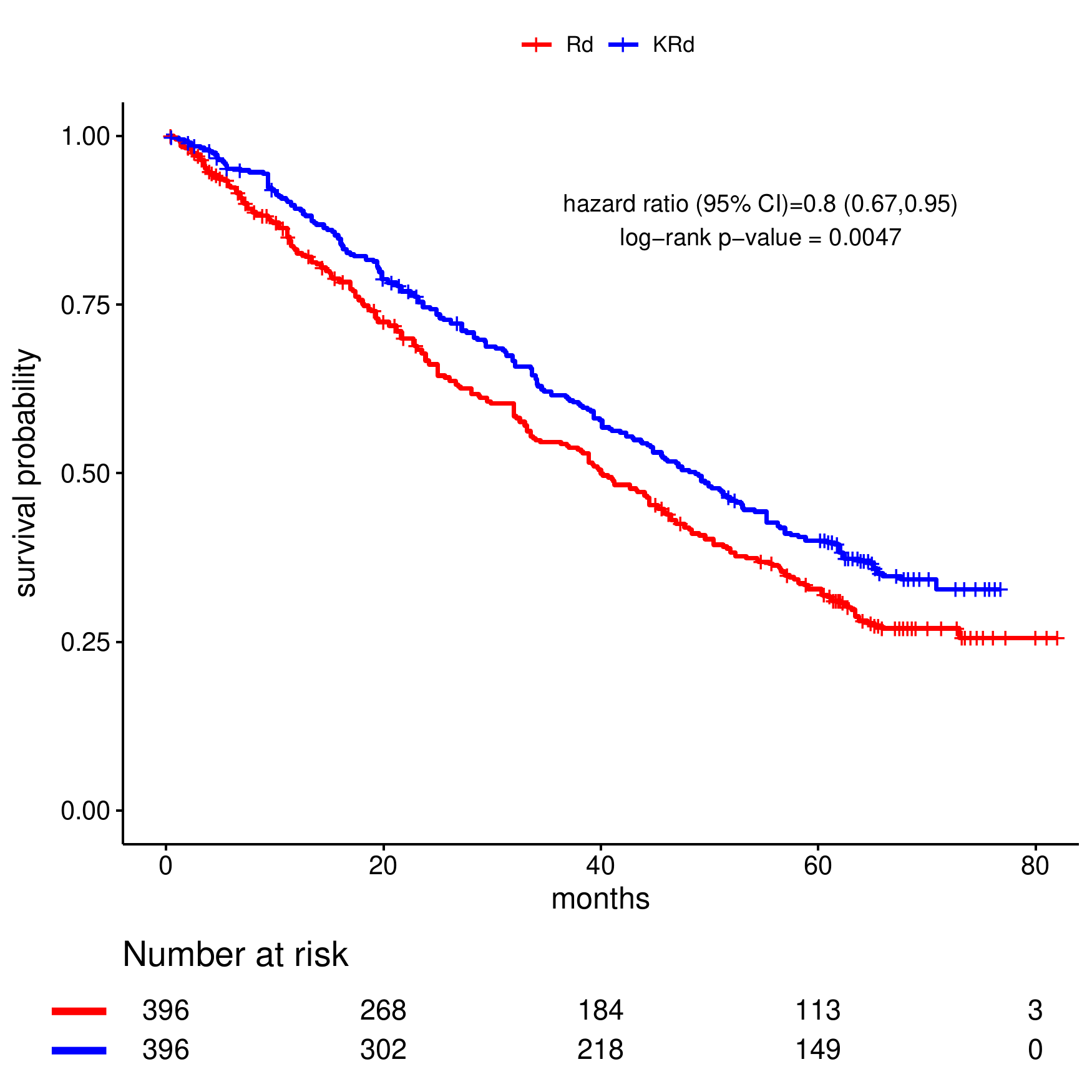}
 \includegraphics[width=0.48\textwidth]{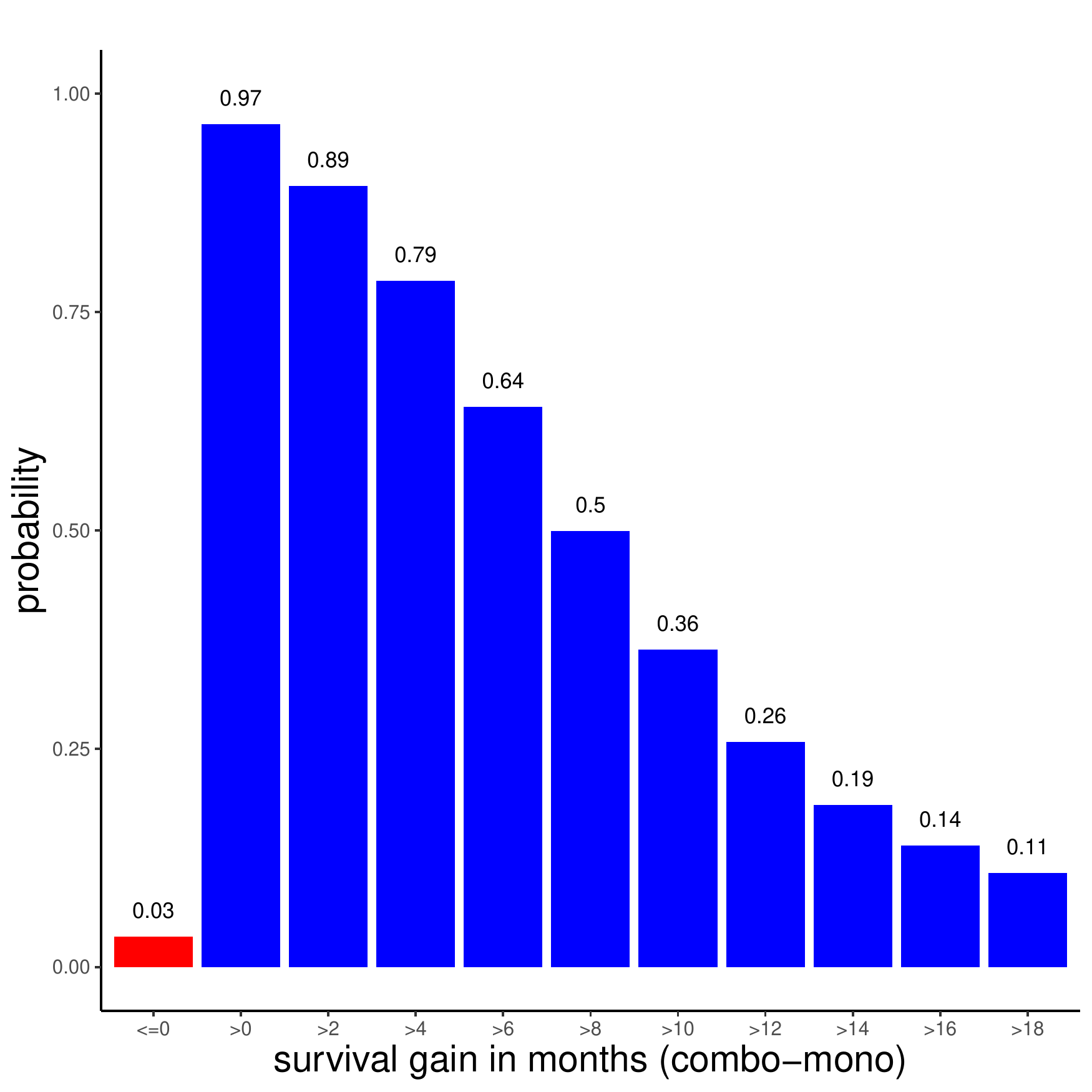}
 \includegraphics[width=0.48\textwidth]{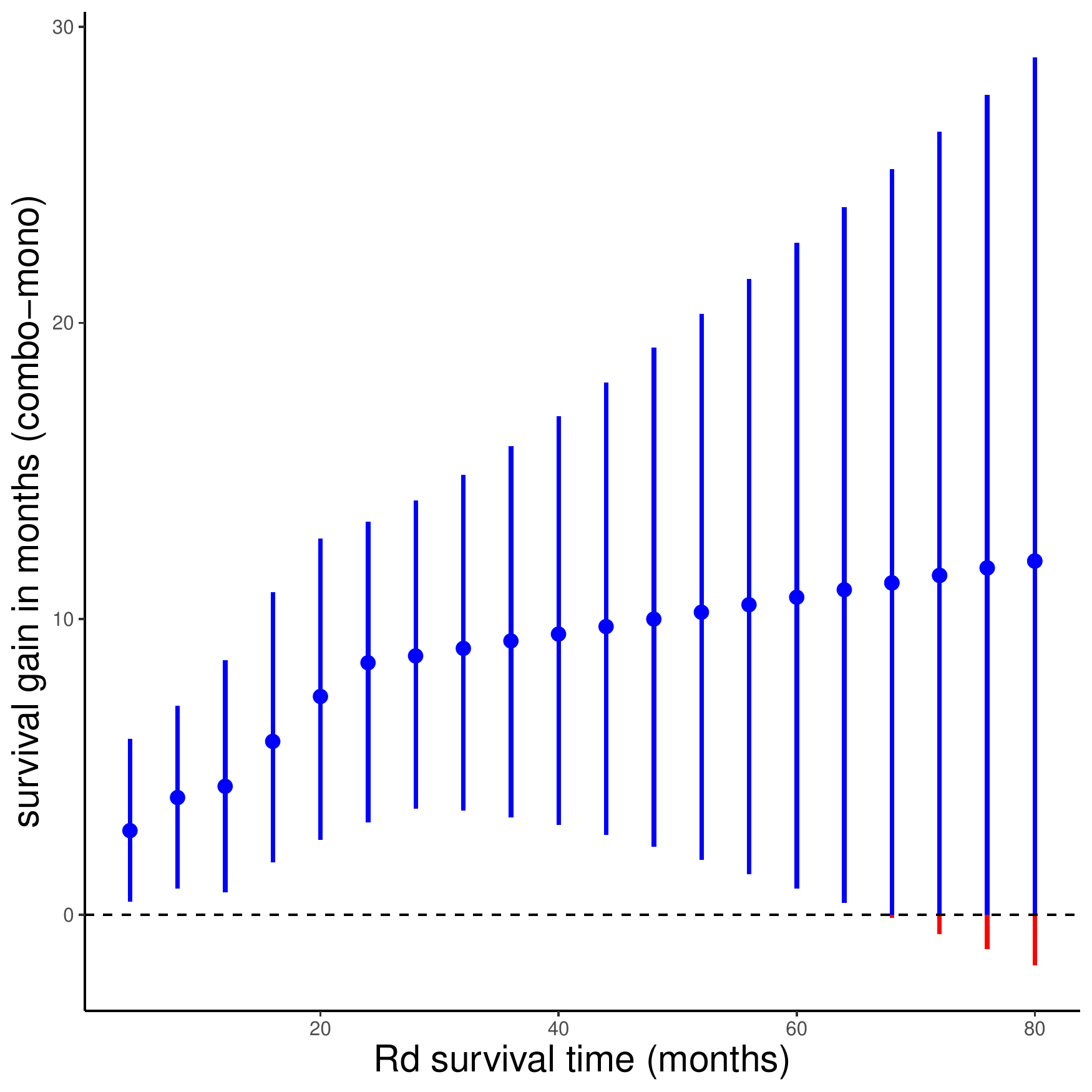}
\end{figure}

%
%
%
%

\end{document}